\documentclass[a4paper,11pt]{article}
\usepackage{pos}
\usepackage{xcolor}
\usepackage{enumitem}
\usepackage{todonotes}
\usepackage{booktabs}
\usepackage{caption}
\usepackage{wrapfig}
\title{Updated comparison of the UHECR energy spectra measured by the Pierre Auger Observatory and the Telescope Array}
\ShortTitle{Updated Comparison of UHECR Spectra from Auger and TA}

\author{Douglas R.~Bergman} 
\author{Toshihiro Fujii} 
\author{Kozo Fujisue}
\author{Keitaro Fujita}
\author{Jihyun Kim}
\author{Diego Ravignani}
\author{Felix Riehn}
\author{Markus Roth}
\author*[a,b]{Francesco~Salamida}
\author{Yoshiki Tsunesada}
\author{Valerio Verzi}

\affiliation[a]{INFN Laboratori Nazionali del Gran Sasso, Assergi (L’Aquila), Italy}
\affiliation[b]{Dipartimento di Scienze Fisiche e Chimiche, Universit\`a dell’Aquila, L’Aquila, Italy}

\onbehalf{for the Pierre Auger Collaboration$^c$ and the Telescope Array Collaboration$^d$}

\affiliation[c]{Observatorio Pierre Auger, Av.\ San Mart{\'\i}n Norte 304, 5613 Malarg\"ue, Argentina}
\affiliation[d]{Telescope Array Project, 201 James Fletcher Bldg, 115 S. 1400 East, Salt Lake City, UT 84112-0830, USA}

\emailAdd{spokespersons@auger.org, TelescopeArray-spokespersons@cosmic.utah.edu}

\abstract{The Pierre Auger and Telescope Array joint Working Group on the UHECR energy spectrum was established in 2012 to analyze energy scale uncertainties in both experiments and to investigate their systematic differences, particularly in the spectral shape of the flux measurements. Previous studies have indeed shown that, within systematic uncertainties, the energy spectra measured by the two observatories are consistent below~$10\,\mathrm{EeV}$. However, at higher energies, a significant difference remains. In this work, we re-examine this discrepancy in greater detail and explore its possible origins. We consider systematic and statistical uncertainties, including the conversion from directly measured observables to energy and the calculation of exposures. We present an updated energy scale comparison between the two experiments and updated flux measurements in the common declination band.}

\ConferenceLogo{PoS_ICRC2025_logo}

\FullConference{
39th International Cosmic Ray Conference (ICRC2025)\\
15 -- 24 July, 2025\\
Geneva, Switzerland}

\begin{document}
\maketitle

\section{Introduction}
\label{sec:intro}

Ultra-high-energy cosmic rays (UHECRs) are atomic nuclei reaching Earth with energies greater than~$10^{18}\,\mathrm{eV}$, the highest observed in nature. Their energy spectrum, describing the differential flux as a function of energy, is a key observable to investigate the sources, composition, and propagation mechanisms of these particles. Due to their extremely low flux, less than one particle per square kilometer per century above $10^{20}\,\mathrm{eV}$, such measurements require detectors with large exposures and long-term operation.

The two largest observatories dedicated to UHECRs are the Pierre Auger Observatory (Auger)~\cite{AugerDet} and the Telescope Array (TA)~\cite{TASD,TAFD}, operating respectively in the Southern and Northern Hemispheres for nearly two decades. Auger, located near Malargüe, Argentina ($35.2^\circ$~S), consists of 1\,600~water-Cherenkov detectors deployed over approximately~$3\,000\,\mathrm{km}^2$ on a triangular grid with 1\,500~m~spacing. TA, located near Delta, Utah ($39.3^\circ$~N), features 507~plastic scintillation detectors arranged on a 1.2\,km~square grid covering approximately~$700\,\mathrm{km}^2$. Both experiments are hybrid, combining surface detectors (SD) and fluorescence detectors (FD\@). The FD provides a calorimetric energy measurement used to calibrate the SD-based energy estimators.

The analysis relies on the most recent datasets: from January~1, 2004, to December~31, 2022, for Auger~\cite{AugerArXiv2025}, and from May~11, 2008, to May~10, 2024, for TA~\cite{TASD_spec_ICRC2025}. Compared to previous reports~\cite{PTEP2017,WG-ICRC2023}, both datasets benefit from increased statistics, and the Auger analysis now includes events with large zenith angles. These improvements allow for a more precise comparison of the spectral shapes and for the identification of possible systematic offsets. We present updated results from both collaborations, starting with a comparison of the energy spectra measured over the full declination bands. We then focus on the common sky region observed by both experiments to reduce the impact of possible astrophysical differences related to the different sky regions observed. To interpret the observed differences, we examine the effects of the energy calibration procedures and investigate the role of mass composition by testing in Auger a simulation-based calibration approach, similar in structure to that used by TA, and studying its dependence on the primary mass. In parallel, TA has revisited the energy linearity checks using an extended hybrid event dataset and a data-driven energy estimation approach, and has expanded its Monte Carlo (MC) dataset to include heavier nuclei.

\section{Comparison of the spectra in the full declination bands}

\label{sec:fullband}

\begin{figure}[t!]
    \centering
    \includegraphics[width=0.49\textwidth]{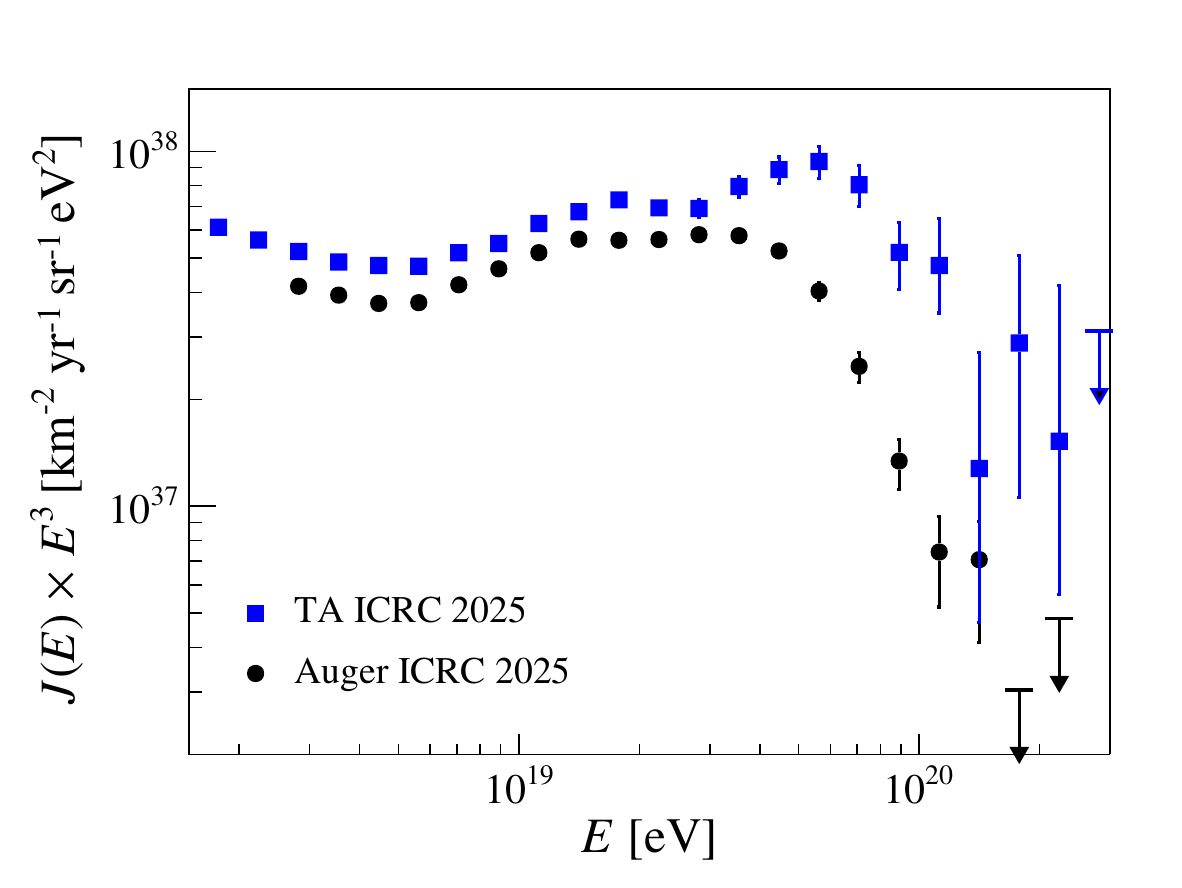}
    \hfil
    \includegraphics[width=0.49\textwidth]{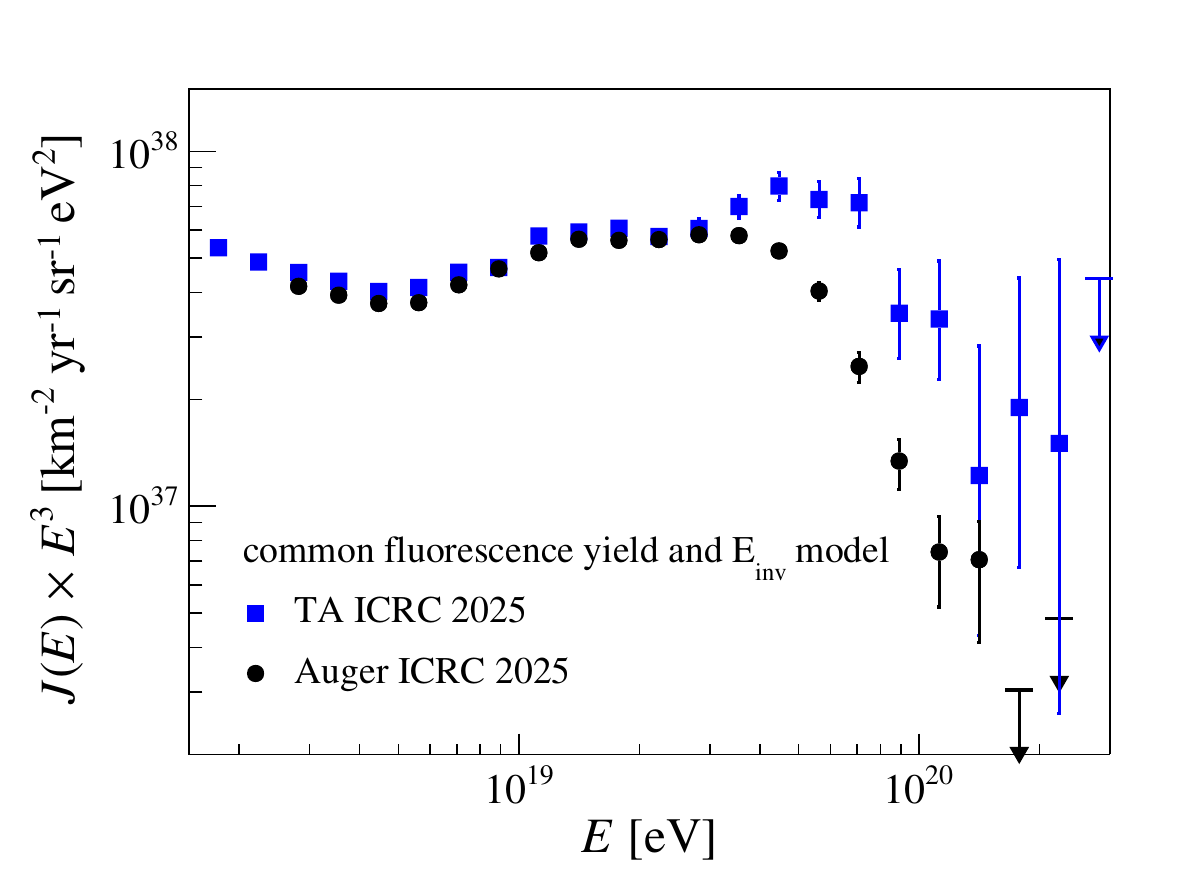}
    \caption{Comparison of the spectra measured in the full declination band. The offset between the two measurements below~$10^{19}\,\mathrm{eV}$ is well described by an overall energy shift of~11.2\% fully compatible within the systematic uncertainties. A better agreement between the spectra is obtained when using the same model for the fluorescence yield and the invisible energy ($E_\text{inv}$) as shown in the right panel. In this configuration, a residual 4\% energy shift remains. This illustrates how calibration choices can account for a substantial part of the observed difference. The interpretation of the comparison at lower energies should be considered with care.}
    \label{fig:spectra_full_band}
\end{figure}

We begin by comparing the energy spectra measured by Auger and TA across their full respective declination ranges. Auger covers declinations from~$-90^\circ$ to~$+44.8^\circ$ (zenith angles up to~$80^\circ$), while TA observes between~$-5.7^\circ$ and~$+90^\circ$ (zenith angles up to~$45^\circ$). Together, they provide full-sky coverage for UHECR studies.

\autoref{fig:spectra_full_band}~(left) shows the two spectra scaled by~$E^3$ to enhance the visibility of spectral features. Below~$\approx 10^{19}\,\mathrm{eV}$, the spectra look very similar, with only an overall offset that can be interpreted with a~11.2\% mismatch between the energy scales of the two observatories, an offset that is fully consistent with the uncertainties in the energy estimation~\cite{WG-ICRC2023}. In the right panel of \autoref{fig:spectra_full_band}, the comparison is repeated by TA while using the same fluorescence yield model (AirFly~\cite{Ave:2012ifa}) and the same treatment of the invisible energy correction~\cite{AugerInvisibleEnergy} as that used by Auger. This common calibration approach reduces the residual energy shift needed to align the spectra to~4.0\%, confirming that a significant part of the observed offset originates from differences in calibration assumptions.

To further quantify the agreement, the spectra were fitted using a sequence of power laws. The fitted parameters, including spectral indices and the energies corresponding to the ankle, the instep, and the suppression, are listed in \autoref{tab:pars}. These fits are performed using the spectra as reconstructed with the standard energy scales of the two experiments. The spectral indices and break energies show overall good agreement between the two experiments. The slopes below and around the ankle ($\gamma_1$, $\gamma_2$) differ by less than~5\% and are compatible within~$2\sigma$, while the position of the ankle~$E_{12}$ agrees within~2\% between Auger and TA. The instep energy~$E_{23}$ is consistent within uncertainties. A significant tension, however, is observed in the suppression energy~$E_{34}$. While Auger finds~$E_{34} = (48 \pm 3)\,\mathrm{EeV}$, TA reports a higher value of~$(68 \pm 5)\,\mathrm{EeV}$, resulting in a relative difference of approximately~42\% and a discrepancy exceeding~$3\sigma$.

\begin{table}[t!]
\centering

\begin{tabular}{lcccc ccc}
\toprule
Parameter & $\gamma_1$ & $\gamma_2$ & $\gamma_3$ & $\gamma_4$ & $\dfrac{E_{12}}{\mathrm{EeV}}$ & $\dfrac{E_{23}}{\mathrm{EeV}}$ & $\dfrac{E_{34}}{\mathrm{EeV}}$ \\[6pt]
\midrule
Auger & $3.26 \pm 0.01$ & $2.51 \pm 0.03$ & $2.99 \pm 0.03$ & $5.3 \pm 0.2$ & $5.1 \pm 0.1$ & $13 \pm 1$ & $48 \pm 3$ \\
TA & $3.28 \pm 0.02$ & $2.62 \pm 0.03$ & $2.82 \pm 0.04$ & $4.7 \pm 0.4$ & $5.0 \pm 0.1$ & $14 \pm 3$ & $68 \pm 5$\\
\bottomrule
\end{tabular}
\caption{\label{tab:pars}Spectral features measured by Auger~\cite{AugerArXiv2025} and TA~\cite{PTEP2017,WG-ICRC2023} in the full declination band. The fit parameters are obtained using the energy spectra reconstructed with the standard energy scale of each experiment.}
\end{table}


\section{Comparison of the spectra in the common declination band}
\label{sec:commonband}

\begin{figure}[t!]
    \centering
    \includegraphics[width=0.49\textwidth]{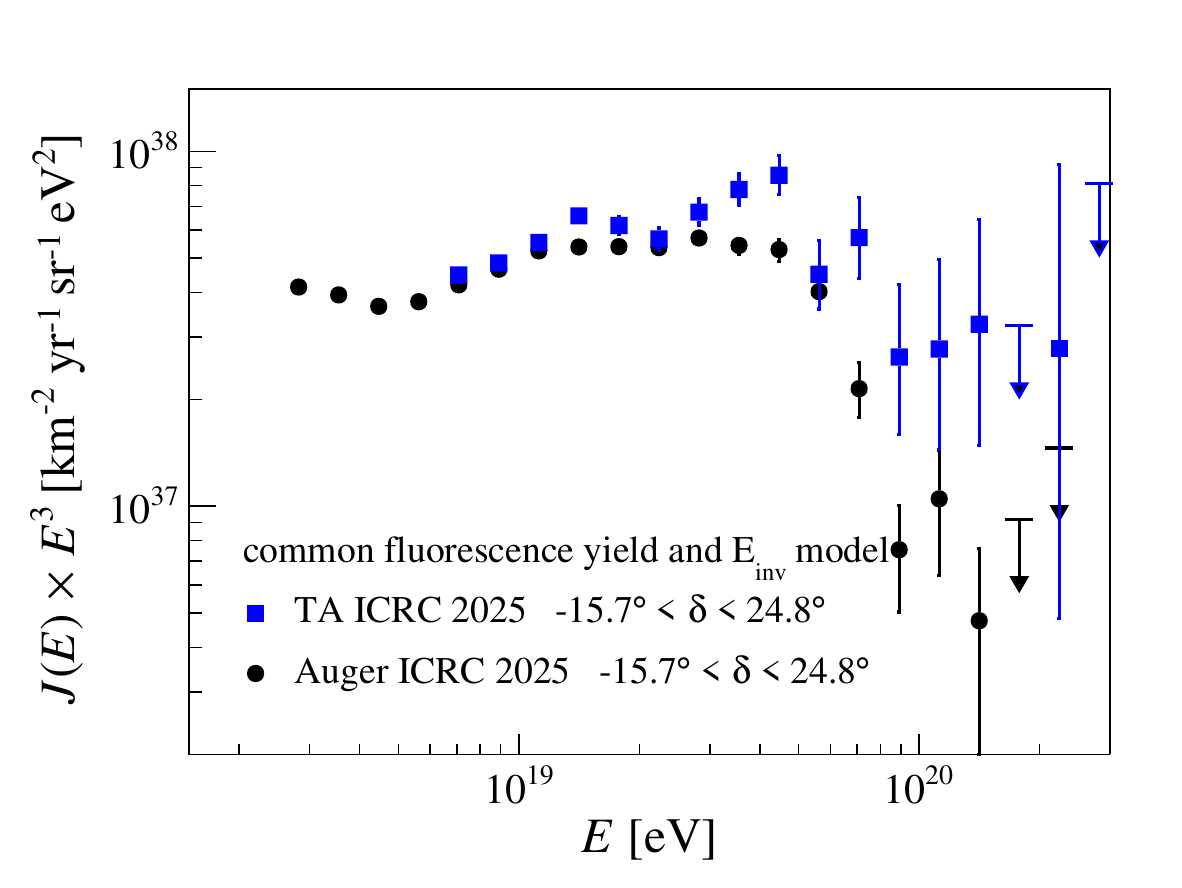}
    \hfil
    \includegraphics[width=0.49\textwidth]{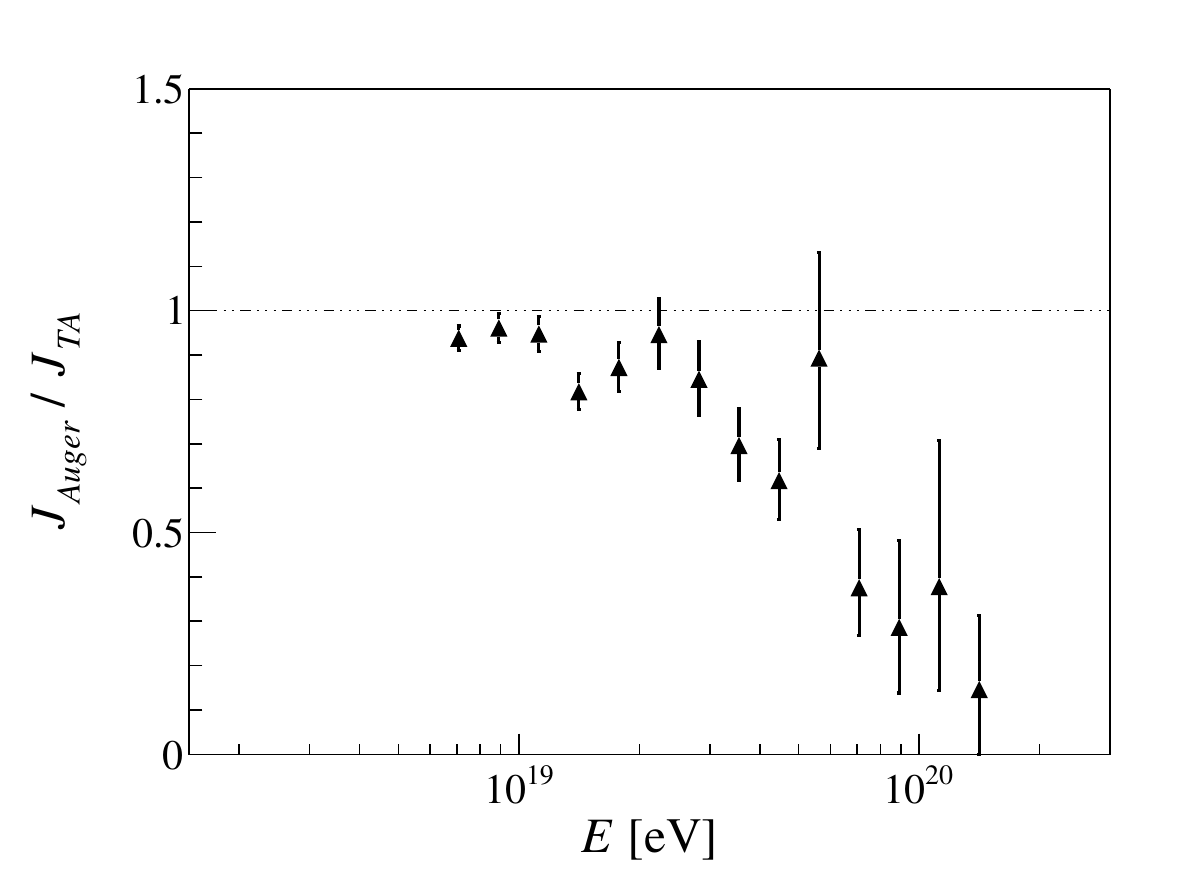}
    \\
    \includegraphics[width=0.49\textwidth]{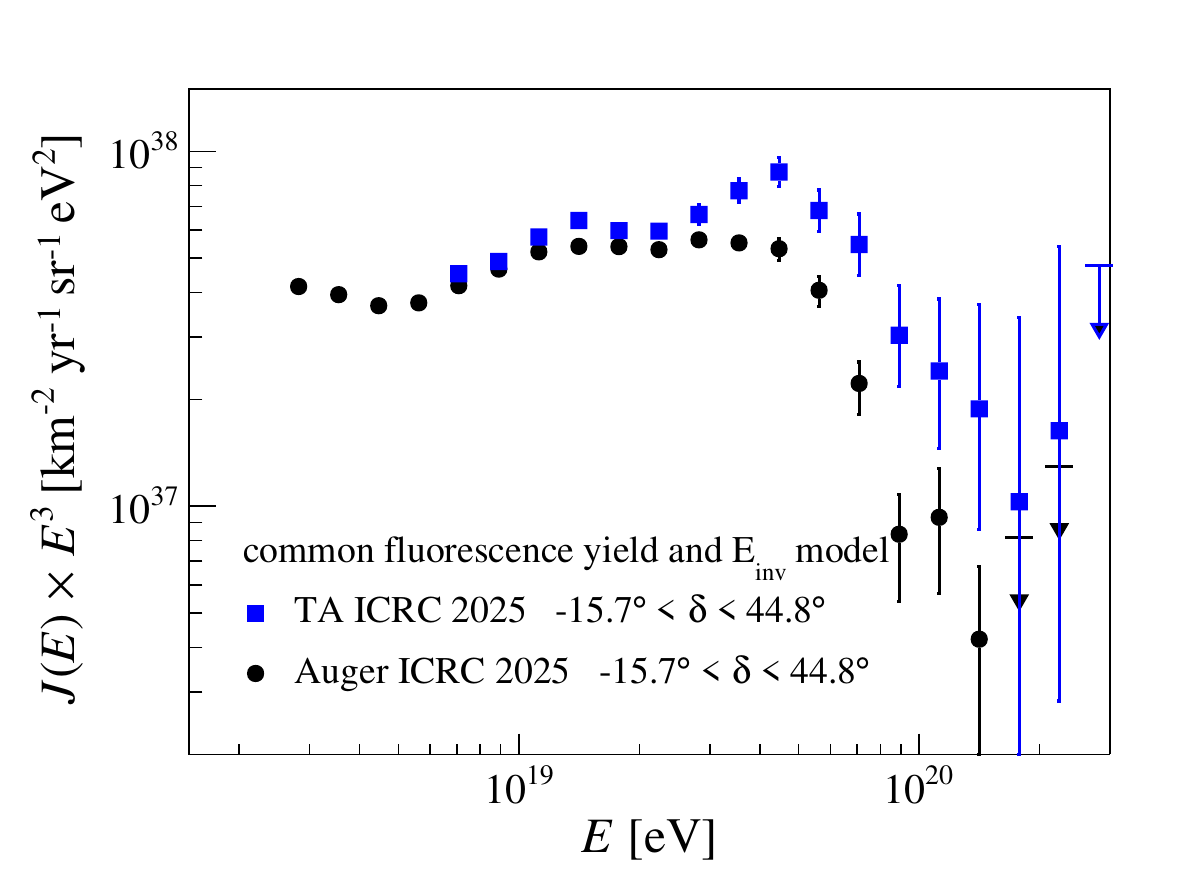}
    \hfil
    \includegraphics[width=0.49\textwidth]{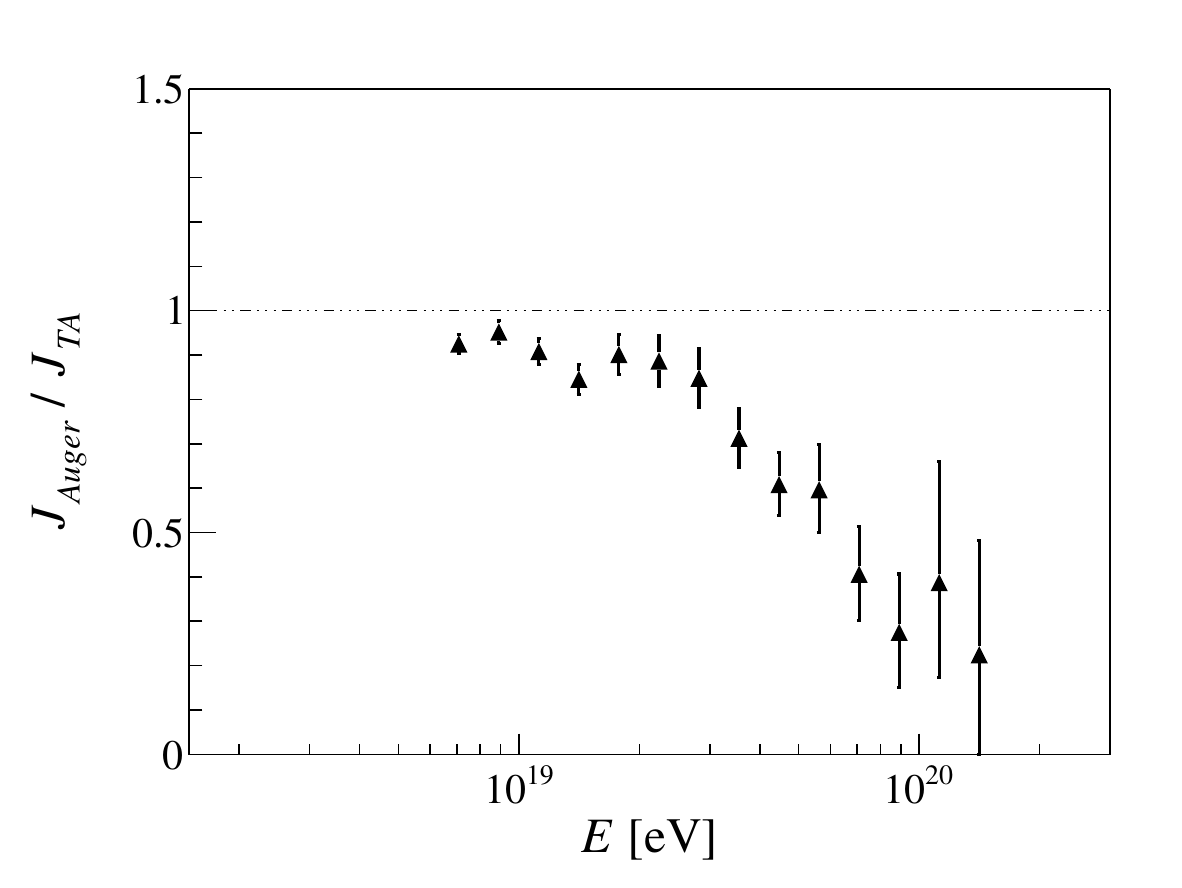}
    \caption{Comparison between the Auger and TA spectra measured in the two common declination bands, the one used in previous reports (upper-left panel) and the enlarged one attainable including the Auger events inclined at large zenith angles (lower-left panel). The spectra are obtained using the same model for the fluorescence yield and the invisible energy ($E_\text{inv}$). The figures in the right panels show the ratio of the spectra shown on the left.} 
   \label{fig:spectra_common_band_unshifted}
\end{figure}

\begin{figure}[t!]
    \centering
    \includegraphics[width=0.49\textwidth]{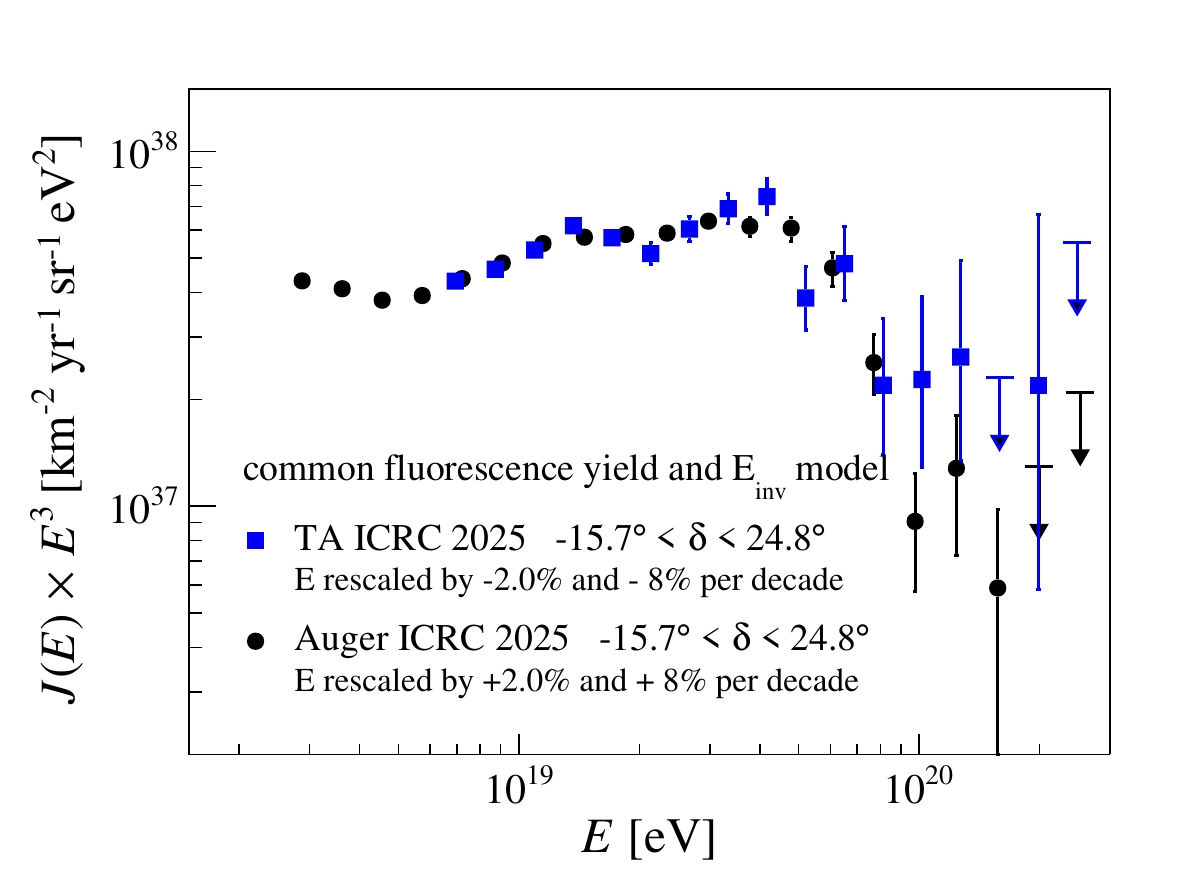}
    \hfil
    \includegraphics[width=0.49\textwidth]{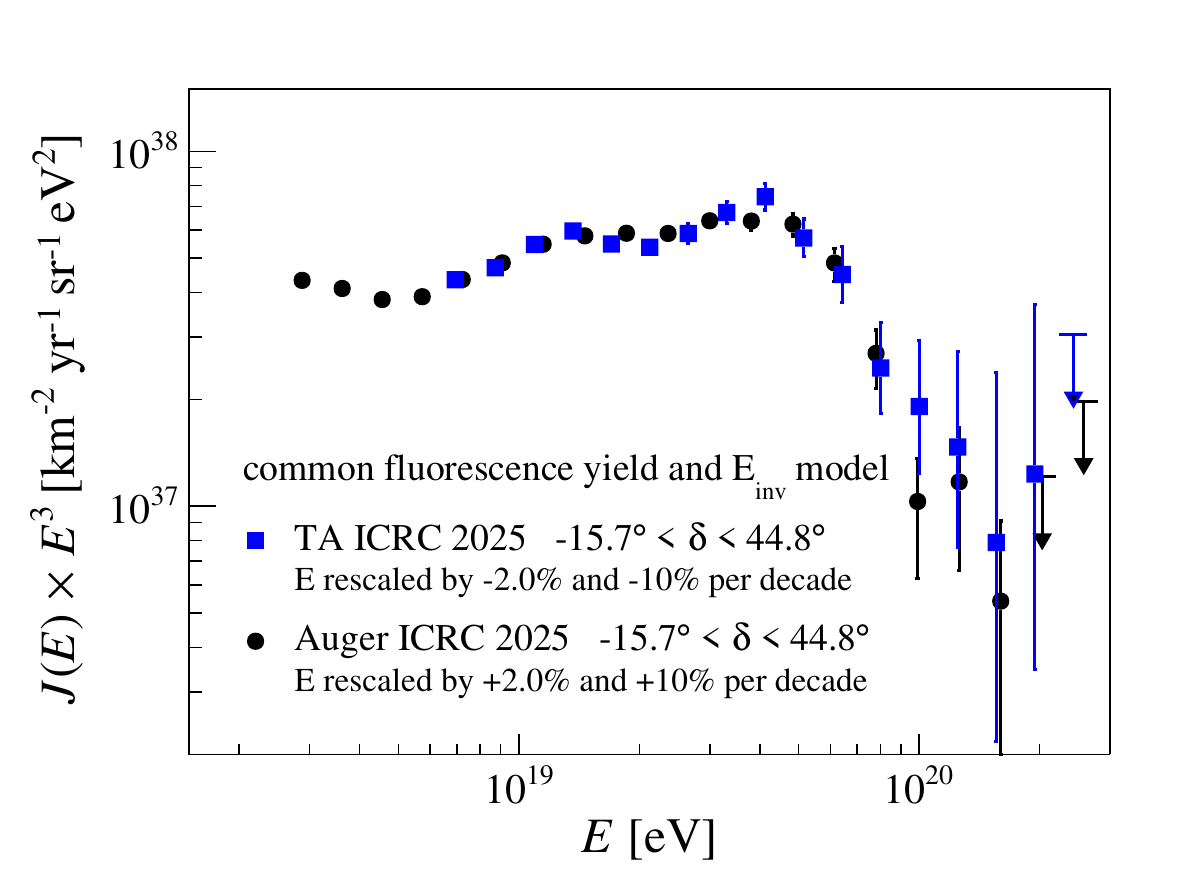}
    \caption{ 
    Auger and TA spectra in the common declination bands after the application of a hypothetical energy-dependent energy shift needed to bring the spectra in agreement. The energy shift is defined as  
     $\Delta E/E = \left[\pm 2.0 \pm \frac{F}{2}(\log_{10}E-19)\right]\%$
     where the constant factor ($\pm 2.0\%$) is the shift determined in the full band and $F$~is 16\% and 20\% for the left and right figure, respectively.}
    \label{fig:spectra_common_band_shifted}
\end{figure}

A central component of the joint analyses performed by the Spectrum Working Group is the comparison of UHECR energy spectra in a common declination band which is observed by both Auger and TA\@. Restricting the comparison to a shared region of the sky reduces potential biases arising from sky-dependent astrophysical effects and allows for a more direct assessment of systematic differences in the reconstructed spectra.

In previous studies, when only Auger events with zenith angles below~$60^\circ$ were used, the common band 
corresponded to declinations between~$-15.7^\circ$ and~$+24.8^\circ$.  We report in \autoref{fig:spectra_common_band_unshifted} (upper panels) an update of the comparison in this 
\textit{original declination band} using the common fluorescence yield and invisible energy as defined in the previous section. The figure on the left shows the two spectra, while their ratio is shown in the right panel. As we can see, there is a significant discrepancy at the highest energies. As shown in the figures of the bottom panels, we repeat the same analysis in an \textit{enlarged common band}, extending the upper declination limit for Auger to~$+44.8^\circ$. This has been achieved by combining vertical and inclined Auger events, including those at large zenith angles, which have been cross-calibrated to ensure a consistent energy scale across the full zenith range~\cite{AugerArXiv2025}.
We note that this band partially includes the localized excesses of events observed by TA, often referred to as the TA Hotspot and the Perseus--Pisces supercluster excess~\cite{TAHotspot2014,TAHotspot2020,TA_anisotropy_UHECR2024}, while no corresponding excess has been reported by Auger in this region~\cite{SGP_Auger2025}. As we can see from the figure in the right panel, the trend is very similar to the one observed in the other band, with a ratio close to one around $10^{19}\,\mathrm{eV}$ and a discrepancy that becomes very significant at the highest energies. The discrepancy observed in the two common bands can be interpreted as a mismatch in the energy scales of the two observatories. As shown in \autoref{fig:spectra_common_band_shifted}, the discrepancy can be eliminated only introducing an energy-dependent energy shift. The shifts applied correspond to the results of fits performed in the full declination band of each experiment.
The shift is estimated as the combination of a $\pm 2.0\%$ factor estimated in the full band analysis below $10^{19}\,\mathrm{eV}$ and an overall shift that amounts to 16\% and 20\% per decade for the \textit{original} and \textit{enlarged} bands, respectively. As shown in previous studies~\cite{SWG-UHECR2016}, the result remains stable when the different directional exposures of the two observatories are properly taken into account, confirming that the observed discrepancy is mainly due to effects of instrumental origin.

In conclusion, the consistency of the required corrections across the same sky region, seen through different experimental coverage, reinforces the hypothesis that the main source of discrepancy may lie in the instrumental and analysis-method domains rather than in differences in the cosmic-ray arrival distribution.

\section{Discussion on systematic uncertainties on energy spectrum measurements}

The investigation of the origin of the energy-dependent differences between the energy spectra measured by Auger and TA has been a long-standing effort within the joint Spectrum Working Group. The evolution of the relative energy scaling between the two experiments has been studied in detail in earlier works~\cite{SWG-UHECR2018, SWG-UHECR2022}, which have shown that the observed differences extend beyond the estimated systematic uncertainties of the two experiments.  
Previous analyses have focused on potential sources such as non-linearity in the surface detector response and uncertainties in the FD energy scale calibration, verifying that possible non-linearity effects are negligible~\cite{SWG-UHECR2022}. In addition, radio-based measurements offer a complementary estimate of the energy scale, independent from the FD, and support the reliability of the current calibration~\cite{radioAugerICRC2025}. In this contribution, we expand the investigation by examining how assumptions on the interaction model and the primary mass composition may impact both the reconstructed energy spectrum and the estimation of the effective exposure.

\subsection{Studies by the Pierre Auger Observatory}

The Auger Collaboration derives the energy spectrum in an almost fully data-driven way~\cite{PhysRevD2020_AugerCIC,AugerArXiv2025}. The SD energy estimator~$S$ is calibrated against the FD energies using a power-law relation~$E_\text{FD} = A S^B$ where the two parameters $A$ and $B$ are fitted to the data. 
Key ingredients of the analysis are also the full trigger efficiency of the array that allows to reduce the calculation of the exposure to a geometrical calculation plus the knowledge of the lifetime of the array, and the estimation of the response function of the array determined through an analysis of the hybrid events.

The fit of the $B$ parameter is important as it allows one to align the SD energies to the FD ones over the full energy range of the measurements, therefore avoiding the bias that likely would appear as a consequence of the change of the mass composition with energy~\cite{AugerMass_ICRC2025}. To better understand this point  
a TA-style lookup table approach, based on MC derived energy estimators, was applied in the energy reconstruction of Auger SD events with zenith angles less than~$60^\circ$. As can be seen in \autoref{fig:AugerMC_TAenRatio} (left panel), using fixed primary assumptions such as pure proton or pure iron, 
the SD/FD energy ratio shows a significant energy-dependent drift. It is then clear that in order to attain a good alignment of the SD to the FD energies it is necessary to fit the $B$ parameter. The fact that the value of the fitted $B$ parameter would be less than 1 suggests that the mass composition evolves from lighter to heavier nuclei because the MC predicts higher energies for lighter primaries (a feature common to all hadronic interaction models). To further explain this point, in the left panel of \autoref{fig:AugerMC_TAenRatio}
we show the results of an analysis in which the lookup table is built taking into account the evolution of the mass composition with energy i.e. AugerMix~\cite{MWG-ICRC2023}. As we can see, in this case the ratio~$E_\text{SD}/E_\text{FD}$ doesn't show any drift, which means that a good energy estimation can be attained fitting only the $A$ parameter leaving $B$ fixed to 1. This internal consistency supports the robustness of the Auger spectrum measurements, which have been shown to be stable across different declination ranges~\cite{AugerArXiv2025}.
 

\subsection{Studies by the Telescope Array}

The TA uses thin scintillator detectors that primarily measure the electromagnetic component, which is well simulated by the MC (and is relatively insensitive to the muonic component), hence the data are well represented by MC. 
The TA energy estimation is based on a lookup table built under the assumption of a specific primary type and hadronic interaction model. 
The standard configuration adopts QGSJet~II-03 proton simulations, as a proxy of light composition, which is consistent with the HiRes and TA composition measurements \cite{HiRes_composition, TA_hybrid_composition, TA_SD_composition,MWG-ICRC2023}. The energy scale is  calibrated to the FD energy scale using hybrid events~\cite{TA2013ApJ}. 
To validate this approach, TA performs extensive data/MC comparisons in many low-level shower parameters~\cite{TAdataMC2014}, which confirm the robustness of MC-based procedures for energy estimation, detector resolution, and acceptance.

\begin{figure}[t!]
    \centering
    \includegraphics[width=0.47\textwidth]{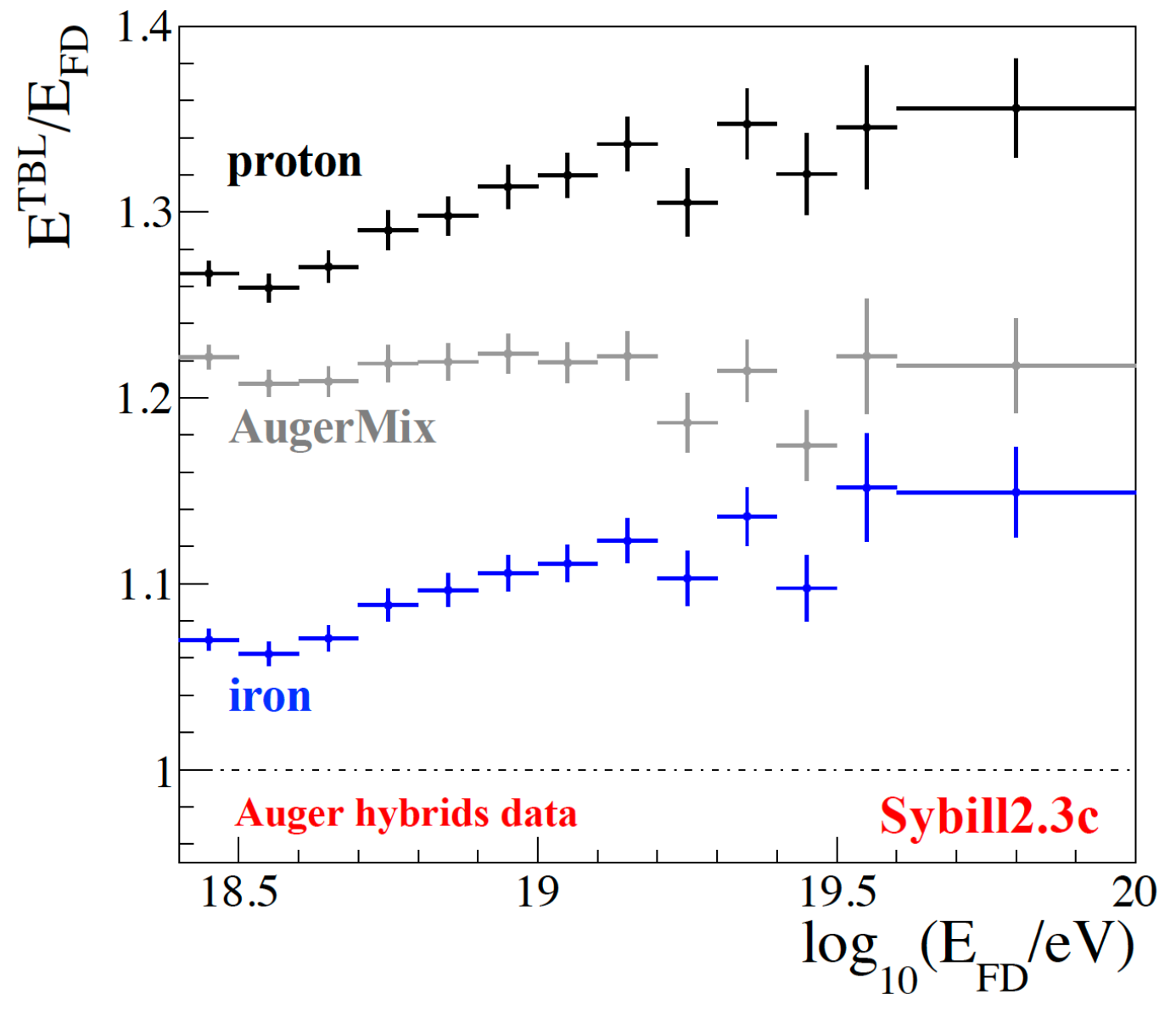}
    \includegraphics[width=0.40\textwidth]{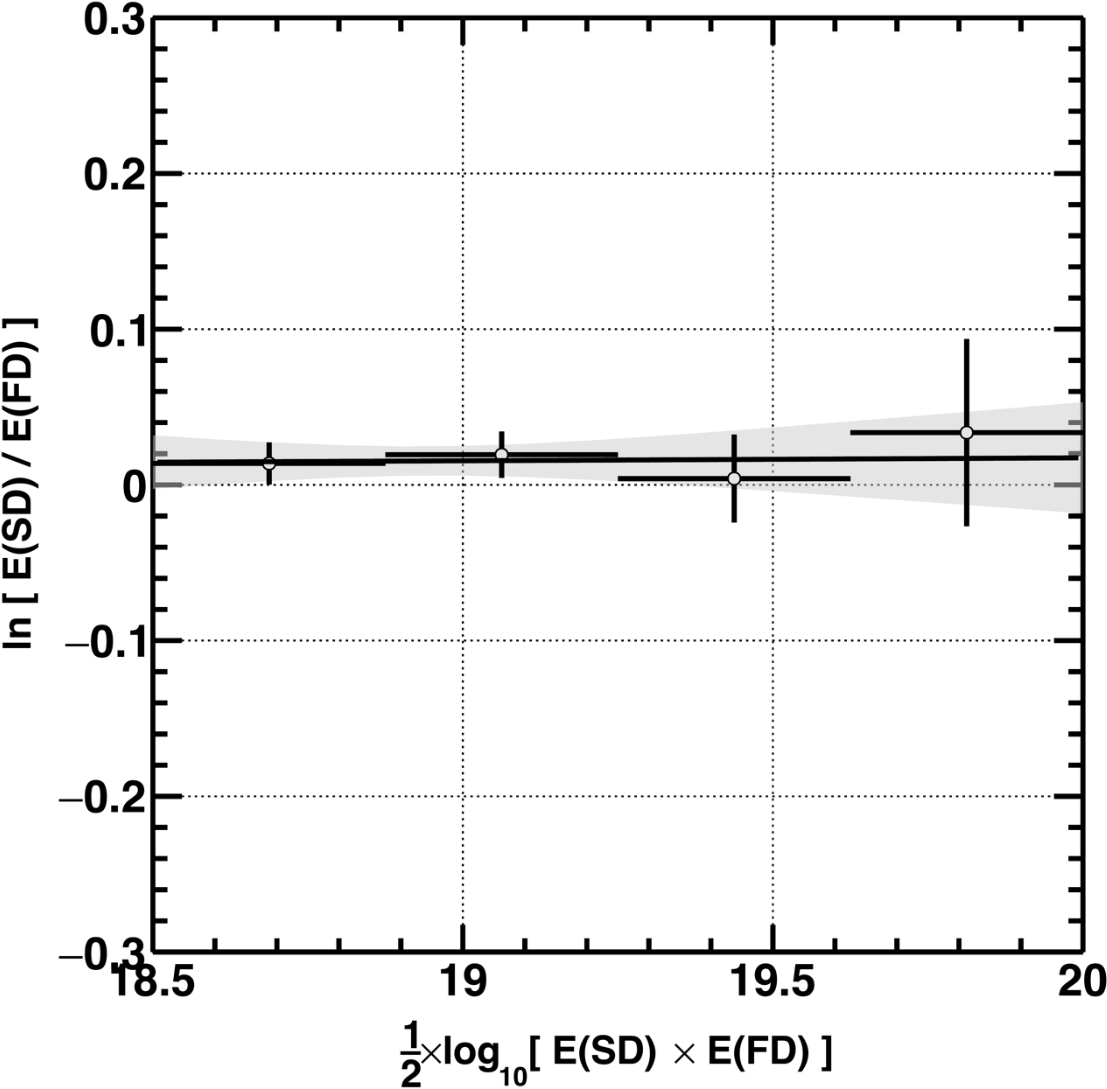}
    \caption{\emph{Left panel}: ratio \(E_\mathrm{TBL}/E_\mathrm{FD}\) for Auger hybrid events using lookup tables built using Sibyll~2.3c under different primary assumptions. A clear energy-dependent drift is observed when using fixed composition assumptions such as pure proton or pure iron. In contrast, adopting the AugerMix composition, which reflects the mass evolution with energy, results in a stable ratio consistent with the standard Auger calibration. \emph{Right panel}: TA's energy ratio plots as a function of energy that represents the SD/FD energy ratio using hybrid events with the standard MC-based SD reconstruction. No energy-dependent non-linearity is observed.
    }
    \label{fig:AugerMC_TAenRatio}
\end{figure}

To probe the linearity of the SD energy reconstruction, TA examined the SD/FD energy ratio using hybrid data. \autoref{fig:AugerMC_TAenRatio} (right) shows that the ratio is stable with energy, ruling out any significant energy-dependent bias. Moreover, the standard TA reconstruction was compared using the data-driven Constant Intensity Cut (CIC) method similar to Auger's and is composition independent: the results of the two reconstructions differ by less than~2\%, which confirms that any composition-dependent bias is minimal~\cite{SWG-UHECR2022, WG-ICRC2023, TA_ICRC2023}. No significant deviation is observed between the two energy estimation methods. 
Further studies investigated a possible composition dependence by generating proton and iron MC\@. The simulations were then reconstructed using both the proton and iron lookup tables. When the thrown composition and lookup table match (e.g., p-p or Fe-Fe), the resulting zenith angle distribution follows the expected~$\sin\theta\cos\theta$ shape.  In contrast, mismatches lead to distortions. 
Since TA sees no such distortion in its data and data/MC comparisons, this is a sign that the MC is a good representation of the data. 

\section{Conclusions}

In this paper we have presented for the first time a detailed comparison of the energy spectra measured by Auger and TA using a common fluorescence yield and invisible energy correction.
The small 4.0\% offset in the energy scales describing the difference among the spectra below $10^{19}\,\mathrm{eV}$ is significantly smaller than the dominant and uncorrelated uncertainties in the absolute calibration of the telescopes 
($\approx10\%$ for both Auger and TA\@). This suggests a good consistency in the FD calibration of the two observatories, a conclusion that remains valid also considering the uncertainties in the exposure (3\% for Auger and well below 10\% for TA above the energy of the ankle). Instead, at the highest energies, above $10^{19}\,\mathrm{eV}$ there is a tension between the two measurements. This discrepancy has been studied in detail in the common declination bands in order to minimize the potential differences in the spectra arising from astrophysical sources.
The study has revealed a significant energy-dependent difference above~$10^{19}\,\mathrm{eV}$  that can be interpreted as an energy shift of 16-20\% per decade.
In order to understand this difference, we have studied the implications of the eventual assumptions on the primary mass composition and hadronic interaction models. The results from the two Collaborations show some disagreement: TA obtains the most precise energy calibration using a QGSJet~II-03 proton lookup table, while Auger, using lookup tables,
concludes that for an unbiased calibration it is necessary to account for the evolution of the mass composition with energy as measured in its data. It is worth noting that the comparison of the two studies is not so straightforward given the different sensitivity of the water Cherenkov and scintillator detectors to muons. 

These results highlight the importance of controlling systematic uncertainties related to the detector response. Future joint studies will be essential to refine the agreement between the two measurements. In particular, the surface scintillator detectors installed on all Auger water-Cherenkov stations as part of the AugerPrime upgrade~\cite{AugerPrime_ICRC2025}, together with the Auger@TA~\cite{AugerAtTA_ICRC2025} and the EarthCARE~\cite{EarthCare_ICRC2025} programs, will offer opportunities to directly compare detection techniques. Finally, a precise understanding of these systematic effects is not only necessary to reconcile the spectra measured by the two experiments, but also to enable the investigation of more fundamental questions. Once instrumental and methodological uncertainties are fully under control, the possibility that part of the observed discrepancies may reflect genuine astrophysical differences can be more robustly addressed.


%


\clearpage
\section*{The Pierre Auger Collaboration}
{\footnotesize\setlength{\baselineskip}{10pt}
\noindent
\begin{wrapfigure}[11]{l}{0.12\linewidth}
\vspace{-4pt}
\includegraphics[width=0.98\linewidth]{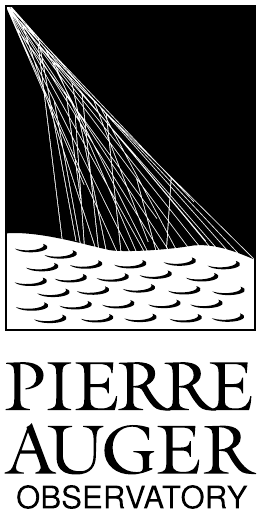}
\end{wrapfigure}
\begin{sloppypar}\noindent
A.~Abdul Halim$^{13}$,
P.~Abreu$^{70}$,
M.~Aglietta$^{53,51}$,
I.~Allekotte$^{1}$,
K.~Almeida Cheminant$^{78,77}$,
A.~Almela$^{7,12}$,
R.~Aloisio$^{44,45}$,
J.~Alvarez-Mu\~niz$^{76}$,
A.~Ambrosone$^{44}$,
J.~Ammerman Yebra$^{76}$
% ...

\end{sloppypar}
\begin{center}
\end{center}

\vspace{1ex}
\begin{description}[labelsep=0.2em,align=right,labelwidth=0.7em,labelindent=0em,leftmargin=2em,noitemsep,before={\renewcommand\makelabel[1]{##1 }}]
\item[$^{1}$] Centro At\'omico Bariloche and Instituto Balseiro (CNEA-UNCuyo-CONICET), San Carlos de Bariloche, Argentina
\item[$^{2}$] Departamento de F\'\i{}sica and Departamento de Ciencias de la Atm\'osfera y los Oc\'eanos, FCEyN, Universidad de Buenos Aires and CONICET, Buenos Aires, Argentina
\item[$^{3}$] IFLP, Universidad Nacional de La Plata and CONICET, La Plata, Argentina
\item[$^{4}$] Instituto de Astronom\'\i{}a y F\'\i{}sica del Espacio (IAFE, CONICET-UBA), Buenos Aires, Argentina
\item[$^{5}$] Instituto de F\'\i{}sica de Rosario (IFIR) -- CONICET/U.N.R.\ and Facultad de Ciencias Bioqu\'\i{}micas y Farmac\'euticas U.N.R., Rosario, Argentina
\item[$^{6}$] Instituto de Tecnolog\'\i{}as en Detecci\'on y Astropart\'\i{}culas (CNEA, CONICET, UNSAM), and Universidad Tecnol\'ogica Nacional -- Facultad Regional Mendoza (CONICET/CNEA), Mendoza, Argentina
\item[$^{7}$] Instituto de Tecnolog\'\i{}as en Detecci\'on y Astropart\'\i{}culas (CNEA, CONICET, UNSAM), Buenos Aires, Argentina
\item[$^{8}$] International Center of Advanced Studies and Instituto de Ciencias F\'\i{}sicas, ECyT-UNSAM and CONICET, Campus Miguelete -- San Mart\'\i{}n, Buenos Aires, Argentina
\item[$^{9}$] Laboratorio Atm\'osfera -- Departamento de Investigaciones en L\'aseres y sus Aplicaciones -- UNIDEF (CITEDEF-CONICET), Argentina
\item[$^{10}$] Observatorio Pierre Auger, Malarg\"ue, Argentina
\item[$^{11}$] Observatorio Pierre Auger and Comisi\'on Nacional de Energ\'\i{}a At\'omica, Malarg\"ue, Argentina
\item[$^{12}$] Universidad Tecnol\'ogica Nacional -- Facultad Regional Buenos Aires, Buenos Aires, Argentina
\item[$^{13}$] University of Adelaide, Adelaide, S.A., Australia
\item[$^{14}$] Universit\'e Libre de Bruxelles (ULB), Brussels, Belgium
\item[$^{15}$] Vrije Universiteit Brussels, Brussels, Belgium
\item[$^{16}$] Centro Brasileiro de Pesquisas Fisicas, Rio de Janeiro, RJ, Brazil
\item[$^{17}$] Centro Federal de Educa\c{c}\~ao Tecnol\'ogica Celso Suckow da Fonseca, Petropolis, Brazil
\item[$^{18}$] Instituto Federal de Educa\c{c}\~ao, Ci\^encia e Tecnologia do Rio de Janeiro (IFRJ), Brazil
\item[$^{19}$] Universidade de S\~ao Paulo, Escola de Engenharia de Lorena, Lorena, SP, Brazil
\item[$^{20}$] Universidade de S\~ao Paulo, Instituto de F\'\i{}sica de S\~ao Carlos, S\~ao Carlos, SP, Brazil
\item[$^{21}$] Universidade de S\~ao Paulo, Instituto de F\'\i{}sica, S\~ao Paulo, SP, Brazil
\item[$^{22}$] Universidade Estadual de Campinas (UNICAMP), IFGW, Campinas, SP, Brazil
\item[$^{23}$] Universidade Estadual de Feira de Santana, Feira de Santana, Brazil
\item[$^{24}$] Universidade Federal de Campina Grande, Centro de Ciencias e Tecnologia, Campina Grande, Brazil
\item[$^{25}$] Universidade Federal do ABC, Santo Andr\'e, SP, Brazil
\item[$^{26}$] Universidade Federal do Paran\'a, Setor Palotina, Palotina, Brazil
\item[$^{27}$] Universidade Federal do Rio de Janeiro, Instituto de F\'\i{}sica, Rio de Janeiro, RJ, Brazil
\item[$^{28}$] Universidad de Medell\'\i{}n, Medell\'\i{}n, Colombia
\item[$^{29}$] Universidad Industrial de Santander, Bucaramanga, Colombia
\item[$^{30}$] Charles University, Faculty of Mathematics and Physics, Institute of Particle and Nuclear Physics, Prague, Czech Republic
\item[$^{31}$] Institute of Physics of the Czech Academy of Sciences, Prague, Czech Republic
\item[$^{32}$] Palacky University, Olomouc, Czech Republic
\item[$^{33}$] CNRS/IN2P3, IJCLab, Universit\'e Paris-Saclay, Orsay, France
\item[$^{34}$] Laboratoire de Physique Nucl\'eaire et de Hautes Energies (LPNHE), Sorbonne Universit\'e, Universit\'e de Paris, CNRS-IN2P3, Paris, France
\item[$^{35}$] Univ.\ Grenoble Alpes, CNRS, Grenoble Institute of Engineering Univ.\ Grenoble Alpes, LPSC-IN2P3, 38000 Grenoble, France
\item[$^{36}$] Universit\'e Paris-Saclay, CNRS/IN2P3, IJCLab, Orsay, France
\item[$^{37}$] Bergische Universit\"at Wuppertal, Department of Physics, Wuppertal, Germany
\item[$^{38}$] Karlsruhe Institute of Technology (KIT), Institute for Experimental Particle Physics, Karlsruhe, Germany
\item[$^{39}$] Karlsruhe Institute of Technology (KIT), Institut f\"ur Prozessdatenverarbeitung und Elektronik, Karlsruhe, Germany
\item[$^{40}$] Karlsruhe Institute of Technology (KIT), Institute for Astroparticle Physics, Karlsruhe, Germany
\item[$^{41}$] RWTH Aachen University, III.\ Physikalisches Institut A, Aachen, Germany
\item[$^{42}$] Universit\"at Hamburg, II.\ Institut f\"ur Theoretische Physik, Hamburg, Germany
\item[$^{43}$] Universit\"at Siegen, Department Physik -- Experimentelle Teilchenphysik, Siegen, Germany
\item[$^{44}$] Gran Sasso Science Institute, L'Aquila, Italy
\item[$^{45}$] INFN Laboratori Nazionali del Gran Sasso, Assergi (L'Aquila), Italy
\item[$^{46}$] INFN, Sezione di Catania, Catania, Italy
\item[$^{47}$] INFN, Sezione di Lecce, Lecce, Italy
\item[$^{48}$] INFN, Sezione di Milano, Milano, Italy
\item[$^{49}$] INFN, Sezione di Napoli, Napoli, Italy
\item[$^{50}$] INFN, Sezione di Roma ``Tor Vergata'', Roma, Italy
\item[$^{51}$] INFN, Sezione di Torino, Torino, Italy
\item[$^{52}$] Istituto di Astrofisica Spaziale e Fisica Cosmica di Palermo (INAF), Palermo, Italy
\item[$^{53}$] Osservatorio Astrofisico di Torino (INAF), Torino, Italy
\item[$^{54}$] Politecnico di Milano, Dipartimento di Scienze e Tecnologie Aerospaziali , Milano, Italy
\item[$^{55}$] Universit\`a del Salento, Dipartimento di Matematica e Fisica ``E.\ De Giorgi'', Lecce, Italy
\item[$^{56}$] Universit\`a dell'Aquila, Dipartimento di Scienze Fisiche e Chimiche, L'Aquila, Italy
\item[$^{57}$] Universit\`a di Catania, Dipartimento di Fisica e Astronomia ``Ettore Majorana``, Catania, Italy
\item[$^{58}$] Universit\`a di Milano, Dipartimento di Fisica, Milano, Italy
\item[$^{59}$] Universit\`a di Napoli ``Federico II'', Dipartimento di Fisica ``Ettore Pancini'', Napoli, Italy
\item[$^{60}$] Universit\`a di Palermo, Dipartimento di Fisica e Chimica ''E.\ Segr\`e'', Palermo, Italy
\item[$^{61}$] Universit\`a di Roma ``Tor Vergata'', Dipartimento di Fisica, Roma, Italy
\item[$^{62}$] Universit\`a Torino, Dipartimento di Fisica, Torino, Italy
\item[$^{63}$] Benem\'erita Universidad Aut\'onoma de Puebla, Puebla, M\'exico
\item[$^{64}$] Unidad Profesional Interdisciplinaria en Ingenier\'\i{}a y Tecnolog\'\i{}as Avanzadas del Instituto Polit\'ecnico Nacional (UPIITA-IPN), M\'exico, D.F., M\'exico
\item[$^{65}$] Universidad Aut\'onoma de Chiapas, Tuxtla Guti\'errez, Chiapas, M\'exico
\item[$^{66}$] Universidad Michoacana de San Nicol\'as de Hidalgo, Morelia, Michoac\'an, M\'exico
\item[$^{67}$] Universidad Nacional Aut\'onoma de M\'exico, M\'exico, D.F., M\'exico
\item[$^{68}$] Institute of Nuclear Physics PAN, Krakow, Poland
\item[$^{69}$] University of \L{}\'od\'z, Faculty of High-Energy Astrophysics,\L{}\'od\'z, Poland
\item[$^{70}$] Laborat\'orio de Instrumenta\c{c}\~ao e F\'\i{}sica Experimental de Part\'\i{}culas -- LIP and Instituto Superior T\'ecnico -- IST, Universidade de Lisboa -- UL, Lisboa, Portugal
\item[$^{71}$] ``Horia Hulubei'' National Institute for Physics and Nuclear Engineering, Bucharest-Magurele, Romania
\item[$^{72}$] Institute of Space Science, Bucharest-Magurele, Romania
\item[$^{73}$] Center for Astrophysics and Cosmology (CAC), University of Nova Gorica, Nova Gorica, Slovenia
\item[$^{74}$] Experimental Particle Physics Department, J.\ Stefan Institute, Ljubljana, Slovenia
\item[$^{75}$] Universidad de Granada and C.A.F.P.E., Granada, Spain
\item[$^{76}$] Instituto Galego de F\'\i{}sica de Altas Enerx\'\i{}as (IGFAE), Universidade de Santiago de Compostela, Santiago de Compostela, Spain
\item[$^{77}$] IMAPP, Radboud University Nijmegen, Nijmegen, The Netherlands
\item[$^{78}$] Nationaal Instituut voor Kernfysica en Hoge Energie Fysica (NIKHEF), Science Park, Amsterdam, The Netherlands
\item[$^{79}$] Stichting Astronomisch Onderzoek in Nederland (ASTRON), Dwingeloo, The Netherlands
\item[$^{80}$] Universiteit van Amsterdam, Faculty of Science, Amsterdam, The Netherlands
\item[$^{81}$] Case Western Reserve University, Cleveland, OH, USA
\item[$^{82}$] Colorado School of Mines, Golden, CO, USA
\item[$^{83}$] Department of Physics and Astronomy, Lehman College, City University of New York, Bronx, NY, USA
\item[$^{84}$] Michigan Technological University, Houghton, MI, USA
\item[$^{85}$] New York University, New York, NY, USA
\item[$^{86}$] University of Chicago, Enrico Fermi Institute, Chicago, IL, USA
\item[$^{87}$] University of Delaware, Department of Physics and Astronomy, Bartol Research Institute, Newark, DE, USA
\item[] -----
\item[$^{a}$] Max-Planck-Institut f\"ur Radioastronomie, Bonn, Germany
\item[$^{b}$] also at Kapteyn Institute, University of Groningen, Groningen, The Netherlands
\item[$^{c}$] Fermi National Accelerator Laboratory, Fermilab, Batavia, IL, USA
\item[$^{d}$] Pennsylvania State University, University Park, PA, USA
\item[$^{e}$] Colorado State University, Fort Collins, CO, USA
\item[$^{f}$] Louisiana State University, Baton Rouge, LA, USA
\item[$^{g}$] now at Graduate School of Science, Osaka Metropolitan University, Osaka, Japan
\item[$^{h}$] Institut universitaire de France (IUF), France
\item[$^{i}$] now at Technische Universit\"at Dortmund and Ruhr-Universit\"at Bochum, Dortmund and Bochum, Germany
\end{description}

\section*{Acknowledgments}

\begin{sloppypar}
The successful installation, commissioning, and operation of the Pierre
Auger Observatory would not have been possible without the strong
commitment and effort from the technical and administrative staff in
Malarg\"ue. We are very grateful to the following agencies and
organizations for financial support:
\end{sloppypar}

\begin{sloppypar}
Argentina -- Comisi\'on Nacional de Energ\'\i{}a At\'omica; Agencia Nacional de
Promoci\'on Cient\'\i{}fica y Tecnol\'ogica (ANPCyT); Consejo Nacional de
Investigaciones Cient\'\i{}ficas y T\'ecnicas (CONICET); Gobierno de la
Provincia de Mendoza; Municipalidad de Malarg\"ue; NDM Holdings and Valle
Las Le\~nas; in gratitude for their continuing cooperation over land
access; Australia -- the Australian Research Council; Belgium -- Fonds
de la Recherche Scientifique (FNRS); Research Foundation Flanders (FWO),
Marie Curie Action of the European Union Grant No.~101107047; Brazil --
Conselho Nacional de Desenvolvimento Cient\'\i{}fico e Tecnol\'ogico (CNPq);
Financiadora de Estudos e Projetos (FINEP); Funda\c{c}\~ao de Amparo \`a
Pesquisa do Estado de Rio de Janeiro (FAPERJ); S\~ao Paulo Research
Foundation (FAPESP) Grants No.~2019/10151-2, No.~2010/07359-6 and
No.~1999/05404-3; Minist\'erio da Ci\^encia, Tecnologia, Inova\c{c}\~oes e
Comunica\c{c}\~oes (MCTIC); Czech Republic -- GACR 24-13049S, CAS LQ100102401,
MEYS LM2023032, CZ.02.1.01/0.0/0.0/16{\textunderscore}013/0001402,
CZ.02.1.01/0.0/0.0/18{\textunderscore}046/0016010 and
CZ.02.1.01/0.0/0.0/17{\textunderscore}049/0008422 and CZ.02.01.01/00/22{\textunderscore}008/0004632;
France -- Centre de Calcul IN2P3/CNRS; Centre National de la Recherche
Scientifique (CNRS); Conseil R\'egional Ile-de-France; D\'epartement
Physique Nucl\'eaire et Corpusculaire (PNC-IN2P3/CNRS); D\'epartement
Sciences de l'Univers (SDU-INSU/CNRS); Institut Lagrange de Paris (ILP)
Grant No.~LABEX ANR-10-LABX-63 within the Investissements d'Avenir
Programme Grant No.~ANR-11-IDEX-0004-02; Germany -- Bundesministerium
f\"ur Bildung und Forschung (BMBF); Deutsche Forschungsgemeinschaft (DFG);
Finanzministerium Baden-W\"urttemberg; Helmholtz Alliance for
Astroparticle Physics (HAP); Helmholtz-Gemeinschaft Deutscher
Forschungszentren (HGF); Ministerium f\"ur Kultur und Wissenschaft des
Landes Nordrhein-Westfalen; Ministerium f\"ur Wissenschaft, Forschung und
Kunst des Landes Baden-W\"urttemberg; Italy -- Istituto Nazionale di
Fisica Nucleare (INFN); Istituto Nazionale di Astrofisica (INAF);
Ministero dell'Universit\`a e della Ricerca (MUR); CETEMPS Center of
Excellence; Ministero degli Affari Esteri (MAE), ICSC Centro Nazionale
di Ricerca in High Performance Computing, Big Data and Quantum
Computing, funded by European Union NextGenerationEU, reference code
CN{\textunderscore}00000013; M\'exico -- Consejo Nacional de Ciencia y Tecnolog\'\i{}a
(CONACYT) No.~167733; Universidad Nacional Aut\'onoma de M\'exico (UNAM);
PAPIIT DGAPA-UNAM; The Netherlands -- Ministry of Education, Culture and
Science; Netherlands Organisation for Scientific Research (NWO); Dutch
national e-infrastructure with the support of SURF Cooperative; Poland
-- Ministry of Education and Science, grants No.~DIR/WK/2018/11 and
2022/WK/12; National Science Centre, grants No.~2016/22/M/ST9/00198,
2016/23/B/ST9/01635, 2020/39/B/ST9/01398, and 2022/45/B/ST9/02163;
Portugal -- Portuguese national funds and FEDER funds within Programa
Operacional Factores de Competitividade through Funda\c{c}\~ao para a Ci\^encia
e a Tecnologia (COMPETE); Romania -- Ministry of Research, Innovation
and Digitization, CNCS-UEFISCDI, contract no.~30N/2023 under Romanian
National Core Program LAPLAS VII, grant no.~PN 23 21 01 02 and project
number PN-III-P1-1.1-TE-2021-0924/TE57/2022, within PNCDI III; Slovenia
-- Slovenian Research Agency, grants P1-0031, P1-0385, I0-0033, N1-0111;
Spain -- Ministerio de Ciencia e Innovaci\'on/Agencia Estatal de
Investigaci\'on (PID2019-105544GB-I00, PID2022-140510NB-I00 and
RYC2019-027017-I), Xunta de Galicia (CIGUS Network of Research Centers,
Consolidaci\'on 2021 GRC GI-2033, ED431C-2021/22 and ED431F-2022/15),
Junta de Andaluc\'\i{}a (SOMM17/6104/UGR and P18-FR-4314), and the European
Union (Marie Sklodowska-Curie 101065027 and ERDF); USA -- Department of
Energy, Contracts No.~DE-AC02-07CH11359, No.~DE-FR02-04ER41300,
No.~DE-FG02-99ER41107 and No.~DE-SC0011689; National Science Foundation,
Grant No.~0450696, and NSF-2013199; The Grainger Foundation; Marie
Curie-IRSES/EPLANET; European Particle Physics Latin American Network;
and UNESCO.
The authors gratefully acknowledge the
computing time provided on the high-performance computer HoreKa by the National High-Performance Computing
Center at KIT (NHR@KIT). This center is jointly supported by the Federal Ministry of Education and Research and
the Ministry of Science, Research and the Arts of Baden-Württemberg, as part of the National High-Performance
Computing (NHR) joint funding program. HoreKa is partly funded by the German Research Foundation.
\end{sloppypar}

}

\clearpage
\section*{The Telescope Array Collaboration}
{\footnotesize\setlength{\baselineskip}{10pt}
\noindent
\begin{sloppypar}\noindent




\makeatletter
\newcommand{\ssymbol}[1]{^{\@fnsymbol{#1}}}
\makeatother
R.U.~Abbasi$^{1}$,
T.~Abu-Zayyad$^{1,2}$,
M.~Allen$^{2}$,
J.W.~Belz$^{2}$,
D.R.~Bergman$^{2}$,
F.~Bradfield$^{3}$,
I.~Buckland$^{2}$,
W.~Campbell$^{2}$,
B.G.~Cheon$^{4}$,
K.~Endo$^{3}$
% ...
\bigskip
\par\noindent
{\footnotesize\it
$^{1}$ Department of Physics, Loyola University-Chicago, Chicago, Illinois 60660, USA \\
$^{2}$ High Energy Astrophysics Institute and Department of Physics and Astronomy, University of Utah, Salt Lake City, Utah 84112-0830, USA \\
$^{3}$ Graduate School of Science, Osaka Metropolitan University, Sugimoto, Sumiyoshi, Osaka 558-8585, Japan \\
$^{4}$ Department of Physics and The Research Institute of Natural Science, Hanyang University, Seongdong-gu, Seoul 426-791, Korea \\
% ...

\let\thefootnote\relax\footnote{$\ssymbol{1}$ Presently at: Purple Mountain Observatory, Nanjing 210023, China}
\let\thefootnote\relax\footnote{$\ssymbol{2}$ Presently at: Physics Department, Brookhaven National Laboratory, Upton, NY 11973, USA}
\let\thefootnote\relax\footnote{$\ssymbol{3}$ Presently at: Korea Institute of Geoscience and Mineral Resources, Daejeon, 34132, Korea}
\let\thefootnote\relax\footnote{$\ssymbol{4}$ Presently at: NASA Marshall Space Flight Center, Huntsville, Alabama 35812, USA}
\addtocounter{footnote}{-1}\let\thefootnote\svthefootnote
}
\par\noindent

\end{sloppypar}
\begin{center}
\end{center}





﻿\section*{Acknowledgements}

The Telescope Array experiment is supported by the Japan Society for
the Promotion of Science(JSPS) through
Grants-in-Aid
for Priority Area
431,
for Specially Promoted Research
JP21000002,
for Scientific  Research (S)
JP19104006,
for Specially Promoted Research
JP15H05693,
for Scientific  Research (S)
JP19H05607,
for Scientific  Research (S)
JP15H05741,
for Science Research (A)
JP18H03705,
for Young Scientists (A)
JPH26707011,
and for Fostering Joint International Research (B)
JP19KK0074,
by the joint research program of the Institute for Cosmic Ray Research (ICRR), The University of Tokyo;
by the Pioneering Program of RIKEN for the Evolution of Matter in the Universe (r-EMU);
by the U.S. National Science Foundation awards
PHY-1806797, PHY-2012934, PHY-2112904, PHY-2209583, PHY-2209584, and PHY-2310163, as well as AGS-1613260, AGS-1844306, and AGS-2112709;
by the National Research Foundation of Korea
(2017K1A4A3015188, 2020R1A2C1008230, and RS-2025-00556637) ;
by the Ministry of Science and Higher Education of the Russian Federation under the contract 075-15-2024-541, IISN project No. 4.4501.18, by the Belgian Science Policy under IUAP VII/37 (ULB), by National Science Centre in Poland grant 2020/37/B/ST9/01821, by the European Union and Czech Ministry of Education, Youth and Sports through the FORTE project No. CZ.02.01.01/00/22\_008/0004632, and by the Simons Foundation (00001470, NG). This work was partially supported by the grants of the joint research program of the Institute for Space-Earth Environmental Research, Nagoya University and Inter-University Research Program of the Institute for Cosmic Ray Research of University of Tokyo. The foundations of Dr. Ezekiel R. and Edna Wattis Dumke, Willard L. Eccles, and George S. and Dolores Dor\'e Eccles all helped with generous donations. The State of Utah supported the project through its Economic Development Board, and the University of Utah through the Office of the Vice President for Research. The experimental site became available through the cooperation of the Utah School and Institutional Trust Lands Administration (SITLA), U.S. Bureau of Land Management (BLM), and the U.S. Air Force. We appreciate the assistance of the State of Utah and Fillmore offices of the BLM in crafting the Plan of Development for the site.  We thank Patrick A.~Shea who assisted the collaboration with much valuable advice and provided support for the collaboration’s efforts. The people and the officials of Millard County, Utah have been a source of steadfast and warm support for our work which we greatly appreciate. We are indebted to the Millard County Road Department for their efforts to maintain and clear the roads which get us to our sites. We gratefully acknowledge the contribution from the technical staffs of our home institutions. An allocation of computing resources from the Center for High Performance Computing at the University of Utah as well as the Academia Sinica Grid Computing Center (ASGC) is gratefully acknowledged.

}

\end{document}